\documentclass[aps,pre,preprint]{revtex4}

\usepackage{amssymb}
\usepackage{graphicx}
\usepackage{epsf}

\begin{document}

\title{GENERATION AND OBSERVATION OF COHERENT, LONG--LIVED STRUCTURES IN A LASER--PLASMA CHANNEL}

\author{T.~V.~Liseykina}\email{t.liseikina@sns.it}
\thanks{On leave from the Institute for Computational 
Technologies, SD-RAS, Novosibirsk, Russia}
\author{F.~Ceccherini}
\author{F.~Cornolti}
\author{E.~Yu.~Echkina}\thanks{Permanent address: Moscow State University, Moscow, Russia}
\author{A.~Macchi}\email{macchi@df.unipi.it}
\author{F.~Pegoraro}
\thanks{Also at polyLAB, CNR-INFM, Pisa, Italy}
\affiliation{Department of Physics ``E. Fermi'', University of Pisa, Pisa, 
Italy}
\author{M.~Borghesi}
\author{S.~Kar}
\author{L.~Romagnani}
\affiliation{IRCEP and School of Mathematics and Physics, the Queen's University of 
Belfast, Belfast BT7 1NN, UK}
\author{S.~V.~Bulanov}
\affiliation{Kansai Photon Science Institute, Japan Atomic Energy Agency, Kyoto, Japan, and A.M. Prokhorov Institute of General Physics RAS, Moscow, Russia}
\author{O. Willi}
\affiliation{Institut f\(\ddot{u}\)r Laser-und Plasmaphysik, Heinrich-Heine-Universit\(\ddot{a}\)t, D\(\ddot{u}\)usseldorf, Germany}
\author{M. Galimberti}
\affiliation{Intense Laser Irradiation Laboratory, IPCF-CNR, Pisa, Italy}

\begin{abstract}
In recent experiments of laser pulse interaction at relativistic intensities with a low density plasma,
the proton radiography  technique showed evidence of long--lived field structures generated  after the
self--channeling of the laser pulse.
We present 2D particle-in-cell simulations of this interaction regime, where the dynamics of similar
structures has been resolved with high temporal and spatial resolution. An axially symmetrical field pattern,
resembling both soliton--like and vortex structures, has been observed. A study of the physics of such structures and a
comparison with experimental data is in progress.
\end{abstract}

\maketitle

\section{Introduction}
The study of the propagation of intense laser pulses in low density plasmas is relevant to several highly
advanced applications, including electron and ion acceleration \cite{malka05,krushelnik99},
development of X- and \(\gamma\)--ray sources, and fusion neutron production \cite{gamma,neutron}.
It is also of fundamental interest, due to the variety of relativistic and nonlinear
phenomena which arise in the laser-plasma interaction \cite{bulanov01}.

During the interaction of high-intensity laser pulses with matter, strong transient
electric fields result from the charge separation induced either by the ponderomotive
force of the laser or by the instantaneous flow of hot electron currents. Depending on
the interaction conditions, these fields can reach extremely large amplitudes, up to
the TV/m range. Electric fields due to charge separation can drive the expansion of
the ions of the plasmas, leading to production of multi--MeV ion beams in interaction
with thin foils \cite{borghesi06,roth05}, or to Coulomb explosion of plasma channels in interaction with
underdense plasmas.
A major step forward in the detection of such fields has been marked by the
development of the proton imaging and deflectometry techniques which, employing
laser-driven protons as a particle probe, have proven to be an exceptionally
useful tool for the investigation of ultra-fast plasma dynamics.
The proton beams emitted from a laser -- irradiated foil are highly laminar \cite{cowan04} and, for projection purposes,
can be described as emitted from a virtual point--like source located in front of the target \cite{borghesi04}.
A point--like projection imaging scheme is therefore automatically achieved. The unique capability of this technique
to detect electrostatic fields in plasmas has allowed the retrieval of direct information on electric fields arising
through a number of laser--plasma interaction processes. The high temporal resolution, related to the picosecond duration
of the proton burst at the source is fundamental in allowing the detection of highly transient fields following short pulse interaction.

In this paper we present results from some recent experiments as well as the results of the simulations obtained for
the range of numerical parameters close to the experiment.
In these investigations, proton probes have been used to investigate
ultrafast plasma dynamics initiated by high-intensity \((10^{18} \div 10^{19}\) W/cm\(^2)\)
interaction, by detecting the ensuing electric and magnetic fields, as well as
to give evidence of as long--lived field structures generated  after the
self--channeling of the laser pulse. In particular, we
discuss the case of laser pulse propagation through underdense plasmas. Beside providing novel information
on the physics of this interaction regime, the data provides a clear example of
the diagnostic capabilities of probing techniques employing laser-driven multi--MeV
protons.

\section{Experimental setup and simulation approach}
The experiment was carried out at the Rutherford Appleton Laboratory, employing the VULCAN Nd-Glass
laser system \cite{danson}, providing two Chirped Pulse Amplified
(CPA) pulses, with 1.054\(\mu\)m wavelength, synchronized
with picosecond precision. Each of the beams delivered
approximately 30~J on target in 1.3~ps (FWHM) duration. By using off--axis parabolas, the beams were
focused to spots of 10\(\mu\)m (FWHM) achieving peak intensities up to 3\(\times  10^{19}\)W/cm\(^2\).
The short pulses were preceded by an Amplified Spontaneous Emission (ASE)
pedestal of 300~ps duration \cite{gregori}. One of the beams was directed to propagate through He gas from a supersonic nozzle, having
a 2~mm aperture, driven at 50 bar pressure. The interaction was transversely probed by the proton beam
produced from the interaction of the second CPA beam with a flat foil (a 10\(\mu\)m thick Au foil was typically used),
under the point projection imaging scheme (see Fig.~\ref{fig:fig1}).
Even if the targets nominally do not contain hydrogen,
hydrocarbon contaminants that are always present on the target
surfaces ensure a suitable source of protons \cite{borghesi03}. The detector was a multilayered radiochromic film detector placed at distance
\(2\div 3\)~cm from the gas jet. In the condition of the experiment, this provided a multi--frame temporal scan of the interaction for up to
50~ps in a single shot \cite{borghesi02}. The temporal resolution of each frame was typically of order of few picoseconds.

For better understanding of the processes observed in the laboratory experiment particle-in-cell (PIC) simulations were carried out.
The numerical model is based on Vlasov equations for both electron and ions and Maxwell equations for the electromagnetic field.
Schematic drawing of the the laser-plasma interaction geometry for two-dimensional (2D) PIC simulations is shown in Fig.~\ref{fig:fig2}.
The reduction to a 2D geometry is necessary because a fully 3D simulation
with both proper numerical resolution and spatial and
temporal scales equal to the experimental ones is far beyond present-day
supercomputing power. Apart from this unavoidable limitation the simulation
parameters are close to the experimental ones. The simulations were
performed on a Linux cluster at the CINECA supercomputing facility
(Bologna, Italy).

\section{Channeling in underdense plasma}
The data presented in Fig.~\ref{fig:fig3} refers to the early stage of the interaction and clearly show that a charge
displacement channel is produced by the laser pulse. The laser pulse propagates from left to right.

A {\it white} channel with {\it dark} boundaries and {\it bullet} shaped leading
edge is clearly visible in the frames (a) and (b). The channel propagates along the axis with the velocity that is equal to the speed of light
within the experimental error. In the trail of the channel the proton flux distribution around the axis changes qualitatively (see Fig.~\ref{fig:fig3}c),
showing a dark line along the axis. The {\it white} channel indicates the presence of a positively charged region around the laser axis.
In this region the electric field points outward in the radial direction. The central {\it dark} line (labeled as "III" in Fig.~\ref{fig:fig3})
observed at later times in the channel
suggests that at this time the radial electric field must change its sign at some radial position leading to the focusing of the probe towards
the axis.

2D electromagnetic PIC simulations in planar geometry qualitatively reproduce the experimental results. In the simulations the laser
pulse was s--polarized (in the experiment the interaction pulse was polarized along the axis of proton probing), thus \(E_{\mathrm{z}}\)
represents the amplitude of the propagating electromagnetic pulse and \(E_{\mathrm{y}}\) is generated by the space--charge displacement.

The simulation clearly shows the formation of the leading part of the laser pulse. This force generates the radial space--charge
electric field, which points away from the laser propagation axis. In the channel trailing edge one observes two narrow ambipolar fronts,
slowly moving away from the axis. As was shown in Ref.\cite{Kar07} such a radial electric field profile produces a pattern in the proton images
similar to that observed in the experiment.

The essential dynamics of the formation of the charged channel and the
subsequent ion evolution could be reproduced also by using a simple
electrostatic, ponderomotive model in 1D cylindrical geometry
\cite{macchi07}. The simulation of proton deflection in the field patterns
reproduced by such model gave proton images very similar to the experimental
ones, thus validating the approach for a study of ponderomotive ion
acceleration in the channel. Fig.\ref{fig:1D} gives snapshots of the electric
field and density profiles and the ion and electron phase space from a
typical 1D simulation. An interesting effect is the ``echo'' behavior of
the electric field, which almost vanishes at the end of the laser pulse,
to reappear later at a point where the ion fluid, accelerated during the
laser pulses, forms a sharp spike and undergoes
hydrodynamical breaking and lead to electron
heating. The ambipolar field structure can be thus interpreted as the sheath
field generated due to electron heating near the density spike.
More details can be found in Ref.\cite{macchi07}.

\section{Structure formation}
At later times, typically on the order of 6-8 ps, the development of
quasiperiodic modulations inside the channel was observed in the experiment (Fig.~\ref{fig:fig5}). These structures
evolved into circular structures which were observed to decay on hydrodynamic
time scales. These structures are tentatively being interpreted as related to the growth of electromagnetic
solitons inside the channel. "Soliton--like" structures  form in high--intensity interactions
due to trapping of the red-shifted electromagnetic
radiation by the ambient plasma, which behaves as over--dense for them \cite{bulanov99}.
The ponderomotive force associated to the trapped radiation expels the
electrons from the core producing a positively charged sphere, which can
deflect the protons of a probe beam. Solitons (or post--solitons,
i.e. larger structures arising from soliton merging \cite{borghesi02a}) can then be
imprinted over the RCF as a "white" region \cite{hatchett}.

The simulations, al later times,  give evidence of the formation in the lower density region
of a deep channel (due to ponderomotive expulsion of electrons and ions
out of the axis). In the higher density region, a variety of nonlinear
effects such as laser beam breakup into filaments and soliton formation
occurs (Fig.~\ref{fig:fig6}).

In the wake of the laser pulse (see Fig.~{\ref{fig:fig7}) we observe the formation of
"standing" (low frequency and almost zero propagation velocity),
structures on the two sides of the laser-drilled channel,
distributed as two (anti)symmetrical rows with respect to the $x$-axis.
Their generation and dynamics may be due to the excitation of
surface modes localized on the channel walls. In fact, the dispersion
relation of surface waves allow the propagation of low-frequency
electromagnetic radiation from the interaction region down to
the low-density plasma. Here, due to the decreasing density,
the group velocity of surface waves slows down and the
field structures become quasi--stationary.

In the direction perpendicular to the simulation plane
these structures appear to have both a steady magnetic field \(B_z\)
(so that they resemble a vortex row) and a steady electric
field \(E_z\) localized into a density depression (analogous to
zero-frequency "solitons").
From the evolution of the ion density
and of the electrostatic field \((E_x, E_y)\) in the simulation plane
we infer that the late evolution of these structure is due to local
ion acceleration long their "radius".
Frequency analysis of fields and currents gives evidence of the
generation of slowly--varying fields.

\section{Conclusion}

In conclusion, we have reported the direct experimental study of electric field dynamics in a charge displacement channel
produced by the interaction of high intensity laser pulse with an underdense plasma.
The use of laser--driven proton probes as a diagnostic tool has proven crucial in obtaining the information
on the structure and development of these fields with picosecond resolution.
The experimental investigations are supported by massively parallel PIC simulations and simplified
analytical/numerical modeling able to addressing the "long" picosecond timescale.

At long times, both experiment and simulation give evidence of long--lived,  "quasi--periodic" field structures in the plasma channel.
We observe both sub-cycle electromagnetic "solitons" at the laser pulse front, and quasi-periodic
structures on the channel sides in the pulse wake. These latter structures
are reminiscent of both vortex and solitonic structures.
The spatio--temporal evolution and the frequency spectrum
of these structures have been characterized in detail, providing a
rich information for the theoretical interpretation.

\section{Acknowledgments}
The authors acknowledge the support of the Royal Society, the QUB/IRCEP scheme,
the RFBR grant, the Italian Ministry for University and Research (MIUR)
via a PRIN project, and the CNR-INFM-CINECA supercomputing initiative. The authors
would like to thank G. Dudnikova, C.A. Cecchetti, R. Jung, J. Osterholz, L. A. Gizzi, J. Fuchs, A. Schiavi and R. Heathcote
 for the useful discussions and the contributions to the experimental work, 
and the staff at the Central Laser Facility, RAL (UK) for the cooperation.

\newpage

\begin{figure}
\caption{Experimental arrangement.}
\label{fig:fig1}
\end{figure}

\begin{figure}
\caption{Initial configuration and the model density profile (left frame). The electron density profile (right frame)
measured along the propagation axis before the high-intensity interaction is consistent with the neutral density
profile of the gas jet and indicates complete ionization by the ASE prepulse \cite{Kar07}.}
\label{fig:fig2}
\end{figure}

\begin{figure}
\caption{Proton projection images of the interaction region
at different times \cite{Kar07}.
The time labels give the probing time of the
proton propagating along the probe axis. White regions correspond to lower proton flux, meanwhile dark regions correspond to
higher proton flux.
}
\label{fig:fig3}
\end{figure}

\begin{figure}
\caption{Distributions of ions density and \(E_{\mathrm{z}}\) and \(E_{\mathrm{y}}\) electric field components obtained from
PIC simulations at 2~ps  \cite{Kar07}.
The bottom plots show the lineout of the radial electrostatic
field at two positions along the $x$ axis.
}
\label{fig:fig4}
\end{figure}

\begin{figure}
\label{fig:1D}
\caption[]{Simulation results from a 1D electrostatic, ponderomotive
model.
The profiles of electric field $E_r$ (blue, thick line)
and ion density $n_i$ (red, dash-dotted line),
and the phase space distributions
of ions $f_i(r,p_r)$ and electrons $f_e(r,p_r)$
are shown at various times.}
\end{figure}

\begin{figure}
\caption{Proton projection images of the interaction region taken at:
(a) 9 ps and (b) 45 ps after the pulse propagation through the center of the frame\cite{borghesi07}.
}
\label{fig:fig5}
\end{figure}

\begin{figure}
\caption{Distribution of the ion density obtained from PIC simulations at 2.83~ps.}
\label{fig:fig6}
\end{figure}

\begin{figure}
\caption{Distribution of the ion density (top frames), \(B_{\mathrm{z}}\) component
of the magnetic field (middle frames) and \(E_{\mathrm{z}}\) component of the electric field (bottom frames)
obtained from PIC simulations at different times.}
\label{fig:fig7}
\end{figure}


\begin{thebibliography}{99}

\bibitem{malka05} V. Malka et al., Plasma Phys. Control. Fusion, {\bf 47}, B481
(2005) and the references therin.

\bibitem{krushelnik99} K. Krushelnick et al., Phys. Rev. Lett., {\bf 89}, 165004 (1999);
M. S. Wei et al., ibid. 93, 155003 (2004)

\bibitem{gamma} A. Rousse et al., Phys. Rev. Lett. {\bf 93}, 135005 (2004).

\bibitem{neutron} S. Fritzler et al., Phys. Rev. Lett. {\bf 89}, 165004 (2002).

\bibitem{bulanov01} S. V. Bulanov et al., Rev. Plasma Phys. {\bf 22}, 227 (2001) and the references therin.

\bibitem{borghesi06} M. Borghesi et. al., Fusion Science and Technology, {\bf 49}, 412 (2006)

\bibitem{roth05} M. Roth et al., Laser and Particle Beams, {\bf 23}, 95 (2005)

\bibitem{cowan04} T.E. Cowan et al., Phys. Rev. Lett., {\bf 92}, 204801 (2004)

\bibitem{borghesi04} M. Borghesi et al., Phys. Rev. Lett., {\bf 92}, 055003 (2004); Bramnik et al., Laser and Particle Beams, {\bf 2}, 163 (2006)

\bibitem{danson} C. Danson et al., J.Mod.Opt. 45, 1653 (1998).

\bibitem{gregori} G. Gregori (RAL), private communication.

\bibitem{borghesi03} M. Borghesi et al., Rev. Sci. Instrum., 74, 1688 (2003).

\bibitem{borghesi02} M. Borghesi et al., Phys. Plasmas, {\bf 9}, 2214 (2002).

\bibitem{Kar07} S. Kar et al., physics/0701332,
submitted to Phys. Rev. Lett. (2007).

\bibitem{macchi07} A. Macchi et al, physics/0701139.

\bibitem{bulanov99} S. V. Bulanov et al., Phys. Rev. Lett., {\bf 82}, 3440 (1999)

\bibitem{borghesi02a} M. Borghesi et al., Phys. Rev. Lett., {\bf 88}, 135002 (2002)

\bibitem{hatchett} S. Hatchett et al., Phys. Plasmas, {\bf 7}, 2076 (2000); S. Wilks et al., Phys. Plasmas, {\bf 8}, 542 (2001)

\bibitem{borghesi07} M. Borghesi et al, Int. J. Mod. Phys. B {\bf 21} (2007), in press.

\end{thebibliography}
\end{document}